\documentstyle[preprint,aps]{revtex}

\begin{document}

\draft

 \title{Exact Foldy-Wouthuysen transformation for
spin $0$ particle in curved space}

\author{Antonio Accioly$^{\ast}$
and Harold Blas$^{\dag}$}

\address{{\it Instituto de F\'{\i}sica Te\'{o}rica, Universidade Estadual
Paulista,}\\ {\it Rua Pamplona 145, 01405-900  S\~{a}o Paulo, SP,
Brazil}\\ }

\maketitle \vskip 1.5 cm

\begin{abstract}
Up to now, the only known exact Foldy- Wouthuysen transformation
(FWT) in curved space is that concerning Dirac particles coupled
to  static spacetime metrics. Here we construct the exact FWT
related to a real spin-0 particle for the aforementioned
spacetimes. This exact transformation exists independently of the
value of the coupling between the scalar field and gravity.
Moreover, the gravitational Darwin term written for the conformal coupling is one third of the relevant term in the fermionic case.
\end{abstract}



\maketitle

\par \vskip .3in \noindent
\vspace{1 cm}


\newpage

The COW experiment \cite{colella} as well as the Bonse-Wroblewski \cite{bonse}
one not only shed a new light on the physical phenomena in which gravitational and quantum effects are interwoven; they also showed that the aforementioned  phenomena are nomore beyond our reach. The theoretical analyssis concerning these experiments  consisted  simply in inserting the Newtonian gravitational potential into the Schr\"odinger equation. To improve their analysis we need to learn certainly how to obtain an adequate interpretation for relativistic wave equations  in curved space. In other words, we have to acquaint ourselves, with the issue of the gravitational effects on quantum mechanical systems. This can be done by constructing the Foldy-Wouthuysen transformation (FWT) \cite{foldy}, \cite{case}- the keystone of relativistic quantum mechanics- for both bosons and fermions coupled to the space-time metric. However, there are very few known problems  in flat space that admit an exact FWT \cite{eriksen}-\cite{nikitin}. In curved space the situation is quite dramatic since up to now the only known exact FWT is that related to Dirac particles coupled to a static spacetime metric \cite{obukhov}.

Here we address ourselves to the problem of finding the exact FWT for a real spin-0 particle coupled to the static metrics
\begin{eqnarray}
\label{metric}
 ds^2 = V^2 dt^2 - W^2 d {\bf{x}}^2,
\end{eqnarray} 
where
$V=V({\bf{x}})$ and $W=W({\bf{x}})$. For the sake of clarification concerning the interpretation of the relativistic single particle wave mechanics for spin-0 boson, we reproduce a remark made by Feshbach and Villars \cite{feshbach} in the 1950's: ``Although it is well known that the Dirac equation gives within proper limits a relativistic wave-mechanical description of a single electron, we find in the literature the (incorrect!) statement that an analogous formalism does not exist for charged spin-0 particles''.

By the middle of the 1970's, Guertin \cite{guertin} constructed the generalized FWT for any $2(2J+1)$-component Poincar\'e- invariant  Hamiltonian theory that describes free massive spin$-J$ particles and that is subject to the conditions: a) every observable is either Hermitian or pseudo-Hermitian and b) the theory is invariant under certain discrete symmetries.

In our convention the signature is $(+ - - - )$. The
curvature tensor is defined by $R^\alpha_{\;\; \beta \gamma
\delta} = -
\partial_\delta \Gamma^\alpha_{\;\; \beta \gamma} +\ldots$, the
Ricci tensor by $R_{\mu\nu} = R^\alpha_{\;\; \mu\nu \alpha}$, and
the curvature scalar by $R = g^{\mu\nu} R_{\mu\nu}$, where
$g_{\mu\nu}$ is the metric tensor. Natural units are used
throughout.   

Currently, we do not have a standard
theory of massive spinless bosons in curved space. That is not the
case as far as the Dirac fermions are concerned. Therefore, our
first task is to find out how the Klein-Gordon (KG) equation
should be written in the general case of a spacetime with
nonvanishing curvature. Let us then start with the following
scalar field equation
\begin{eqnarray}
\label{equation}
 \left( \Box + m^2 + \lambda R \right) \phi =0 \;\;\; , 
\end{eqnarray}
which is obtained from the action
\begin{eqnarray}
\label{action}
S = \int \frac{1}{2} \sqrt{-g}  \left[ g^{\mu\nu}
\partial_{\mu} \phi \; \partial_{\nu} \phi - \left( m^2 + \lambda
R \right) \phi^2 \right] d^4 x \;\;\; . 
\end{eqnarray}

Note that the coupling between the real scalar field $\phi$
and the gravitational field represented by the term $\lambda R
\phi^2$,where $\lambda$ is a numerical factor and $R$ is the Ricci
scalar, is included as the only possible local scalar coupling of
this sort \cite{birrel}. Here
$$ \Box \equiv g^{\mu\nu} \nabla_{\mu}\nabla_{\nu} =
\frac{1}{\sqrt{-g}}\partial_{\mu} \left( \sqrt{-g}
g^{\mu\nu}\partial_{\nu} \right)\;\;\; . $$

The coupling constant $\lambda$, of course, can have any real
value. This raises a delicate question: Which value of $\lambda$
should we single out? Fortunately, there are some arguments that seems to favour the choice $\lambda\,=\, 1/6$:

i) the equation for the massless scalar field is conformally invariant \cite{birrel}-\cite{chernikov};

ii) under the assumption that a) the scalar field satisfies (\ref{equation}), and b) the field $\phi$ does not violate the equivalence principle, the coupling constant is forced to assume the value $1/6$ \cite{sonego}-\cite{faraoni};
 
iii) the minimal coupling leads to a tachyonic behavior whereas the conformal one ($\lambda\,=\, 1/6$) have a correct quasiclassical limit \cite{grib}

There are other reasons (see e.g. \cite{flachi} and references therein) to justify the presence of the nonminimal term in ( \ref{action}).
 
Here we examine the problem in the context of the exact FWT transformations for spin-0 particles. Let us then concentrate our attention on the curved spacetimes
described by Eq. (\ref{metric}). The Ricci scalar related to this metric
is given by
\begin{eqnarray}
\label{ricci}
R = \frac{2}{W^4} \left(\nabla W \right)^2 -
\frac{2}{V W^3} \nabla V \cdot \nabla W -
\frac{2}{V W^2} \nabla^2 V - \frac{4}{W^3}\nabla^2 W. \;\;\;
\end{eqnarray}

Inserting (\ref{ricci}) into (\ref{equation}), we promptly obtain
\begin{eqnarray}
\label{secondorder}
\ddot{\phi}  - F^2 \nabla^2 \phi - F^2 \nabla \ln
(VW) \cdot \nabla \phi + m^2 V^2 \phi + \lambda R
V^2 \phi =0 \;\;\; , 
\end{eqnarray}
where $F^2 \equiv \frac{V^2}{W^2}$. Here the
differentiation with respect to time is denoted by dots.

In order to bring the equation in hand to Schr$\ddot{{\rm
o}}$dinger form we introduce the two-component formalism for the KG equation
$$ \phi = \phi_1 + \phi_2 \;\;\;, \;\;\; \frac{i}{m} \dot{\phi} =
\phi_1 - \phi_{2} \;\;\; . $$

Accordingly, the KG equation can be written in first-order form
\begin{eqnarray} 
\label{schrodinger1}
i \dot{\Phi} = {\cal{H}} \Phi \;\;\;,
\end{eqnarray}
with the Hamiltonian given by
\begin{equation}
\label{hamiltonian}
{\cal{H}} = \frac{m}{2} \xi^T - \xi \theta \;\;\; ,
\end{equation}
where
$$ \Phi = \left(
\begin{array}{c}
  \phi_1  \\
  \phi_2
\end{array}\right)  \;\;\;, \;\;\;
\xi= \left(
\begin{array}{cc}
  1 & 1 \\
 -1 & -1
\end{array}\right)
$$
and the operator $\theta$ is defined by
\begin{eqnarray}
\nonumber
\theta \equiv \frac{F^2}{2m} \nabla^2 - \frac{F^2}{2m}
\nabla \ln (VW) \cdot \nabla -
\frac{m}{2} V^2 - \frac{\lambda}{2 m}V^2 R \;\;\; . 
\end{eqnarray}

Note that the matrix $\xi$ has the following algebraic
properties
$$ \xi^2 = 0 \;\;\; , \;\;\; \left\{ \xi , \xi^{T} \right\} = 4
\;\;\; .$$

It is worth mentioning that the equations of motion derived from (\ref{schrodinger1}) are invariant under ${\cal{H}}\rightarrow -{\cal{H}}^{*}$ and
$\phi_{1, 2} \rightarrow \pm \phi_{2, 1}$, which implies that in
the two-component description of neutral spin-0 particles the
particle and antiparticle may be identified since the
gravitational interaction does not remove the
particle-antiparticle degeneracy.

The operator $\theta$ is formally self-adjoint \cite{nota1} with respect to an inner product provided the spatial integrations are carried out using the correct measure \cite{fulling}
\begin{equation}
 \left\langle {\theta} \right\rangle = \int \rho \; d^3
{\bf{x}} \psi^{\dagger} {\theta} \psi \;\;\; , 
\end{equation}
where $\rho  \equiv g^{00} \sqrt{-g} = \frac{W^3}{V}\;\;\; .$

However, it is more convenient to write the wave
function so that $\theta$ is Hermitian with respect to the
usual flat space measure. We do this by means of a
transformation
$$ \Phi \longrightarrow \Phi' = f \Phi \;\;\; ,\;\;\;{\theta'}=
f {\theta} f^{-1} ,\,\,\,\,\,\mbox{and} \,\,\,\,\,\, {\cal{H}'}=
f {\cal{H}} f^{-1} \;\;\; , $$
 with $ f \equiv \sqrt{\rho} = V^{-1/2} W^{3/2} \;\;\; . $

Therefore $$ {\cal{H}}' = \frac{m}{2} \xi^{T} - \xi
\theta ' \;\;\; , $$
 where
$$ \theta ' \Phi ' = f \theta f^{-1} \Phi '  \;\;\; . $$

Performing the computation, we then find that $\theta '$ can be
written as
\begin{equation}
\label{tetali}
 \theta ' = - \frac{m}{2} V^2 - \frac{1}{2m}F {\hat{p}}^2 F +
\frac{1}{8m} \nabla F \cdot \nabla F + {\cal{D}}_{\lambda}(V, W) \;\;\; ,
\end{equation}
\noindent where  $ \hat{\bf{p}} = - i \nabla $ denotes
the momentum operator and the last term becomes
\begin{eqnarray}
{\cal{D}}_{\lambda}(V, W)\, \equiv \,  \lambda [ (\frac{1}{2\lambda}-2)\frac{V}{W^2}\nabla^2 V- 2\frac{V}{W^3}\nabla V. \nabla W + (\frac{1}{2\lambda}-4) \frac{V^2}{W^3} \nabla^2 W + 2\frac{V^2}{W^4} (\nabla W)^2 ]
\end{eqnarray}

The fascinating property of the transformed Hamiltonian
${\cal{H}}'$ is that its square,

\begin{equation} 
{{\cal{H}}'}^{\Box} = - \frac{m}{2} \theta ' \left\{ \xi
,\xi ^{T} \right\} = -2 m \theta ' I\;\;\; ,
\end{equation} 
where
$$ I = \left(
\begin{array}{cc}
 1 & 0 \\
 0 & 1
\end{array}\right) \;\;\;.
$$

Note that formally
\begin{eqnarray}
\nonumber
\label{root} 
\sqrt{{\cal{H}'}^{\Box}} = (- 2m \theta ' )^{1/2} \,I^{1/2}.
\end{eqnarray}

Since the square root of the $2$x$2$ identity matrix is not unique the FWT transformation needs an extra diagonalizing transformation to the basis where positive and negative energy eigenstates are decoupled. This process can be made with the help of a nondegenerate matrix $U$ such that \cite{nota3}
\begin{eqnarray}
\nonumber
{\cal{H}}'' & \equiv &  (- 2m \theta ' )^{1/2} U\, I^{1/2}\, U^{-1}\\
\nonumber
& =& (- 2m \theta ' )^{1/2} \eta
\end{eqnarray}  
where
$$ \eta= \left(
\begin{array}{cc}
1 &0 \\
  0 & -1
\end{array} \right).
$$

Accordingly, $\cal{H} \rightarrow {\cal{H}}''$ is the exact FWT for
the KG equation in curved space.

Taking (\ref{tetali}) into account, we arrive at the following expression for
the Hamiltonian squared
\begin{eqnarray}
 {{\cal{H}}'}^{\Box} = m^2 V^2 + F \hat{p}^2 F -\frac{1}{4}
\nabla F \cdot \nabla F + {\cal{D}}_{\lambda}(V, W).
\;\;\;  
\end{eqnarray}

The quasirelativistic Hamiltonian is simply obtained by assuming that $m^2$ is the dominating term. We thus arrive at
\begin{equation}
\label{quasire1}
{\cal{H}}'' \approx  \{ m V + \frac{1}{4m}
\left( W^{-1} {\hat{p}}^2 F + F {\hat{p}}^2 W^{-1} \right) -
\frac{1}{8 m V} \nabla F \cdot \nabla F + \frac{1}{2
m V} {\cal{D}}_{\lambda}(V, W)\}\,\,  \eta \;\;\; .
\end{equation}

Some comments are in order here:

{\it (i)} Notice the appearance of a Darwin-like term  $\frac{1}{2
m} {\cal{D}}_{\lambda}(V, W)$ in the quasirelativistic Hamiltonian (\ref{quasire1}). For  $\lambda = 1/6$ conformal invariance constrains the structure of the Darwin-like term to the form
\begin{equation}
\label{darwin}
 \frac{1}{12 m W} \nabla^2 F. \;\;\; 
\end{equation}
Therefore one obtains 
\begin{equation}
\label{quasire2}
{\cal{H}}'' \approx  \{ m V + \frac{1}{4m}
\left( W^{-1} {\hat{p}}^2 F + F {\hat{p}}^2 W^{-1} \right) -
\frac{1}{8 m V} \nabla F \cdot \nabla F + \frac{1}{12
m W} \nabla^2 F \}\,\, \eta.\;\;\;
\end{equation}

{\it (ii)} (\ref{quasire2}) is identical to the spinless sector found
by Obukhov \cite{obukhov} for the Dirac particle except for the Darwin term
which is one third of the corresponding term in the fermionic
case \cite{nota2}. 

iii) The Darwin term (\ref{darwin}) only exists in the context of the exact FWT if the interaction of the scalar field with gravity is of the conformal type $\lambda=1/6$, while for $\lambda \neq 1/6$ the Darwin term is more complicated.

Some remarks about i) and  iii). It is claimed in the literature that (\ref{equation}) with $\lambda = 1/6$ violates the equivalence principle and leads to the appearance of anomalous R-forces between two ``scalar charged'' particles \cite{lightman}. Grib and Poberii \cite{grib} showed, however, that this is not the case. According to them the conformal coupling leads to a correct quasiclassical limit while the minimal one is responsible for a tachyonic behavior.

To conclude we shall prove that the conformal coupling does not violate  the equivalence principle by making a comparison of the true gravitational coupling with the pure inertial case. To do that, we recall that far from the source the solution of the Einstein equation for a point particle of mass M located at $r=0$, is given by
\begin{eqnarray}
\label{g1}
g_{00}\,\approx\, 1-\frac{2MG}{r}\\
\label{g2}
g_{11}\,=\, g_{22}\,=\,g_{33}\,\approx\,-1-\frac{2MG}{r}
\end{eqnarray}

From (\ref{g1}) and (\ref{g2}) we get immediately

\begin{eqnarray}
\label{v1}
V\,\approx\,1-\frac{MG}{r},\,\,\,\,\,\,\,
W\,\approx\,1+\frac{MG}{r}\\
\label{f1}
\mbox{and}\,\,\,\,\,\,  F\,\approx\, 1- 2 \frac{MG}{r}.
\end{eqnarray}

Inserting (\ref{v1}) and (\ref{f1}) into  (\ref{quasire2}) we obtain the nonrelativistic FW Hamiltonian, namely,
\begin{equation}
\label{f12}
{\cal{H}}'' =\left[ {\atop} m+ m \; {\bf g} \cdot
{\bf x} + \frac{\hat{{\bf p}}^2}{2m} + \frac{3}{2m} \hat{{\bf{p}}} \cdot({\bf g} \cdot {\bf x}) \hat{{\bf{p}}} \right] \eta, \;\;\;
\end{equation}
where $ {\bf g} = - GM \frac{{\bf r}}{r^3}$. On the other hand,  in the case of the flat
Minkowski space in accelerated frame,
\begin{eqnarray}
\nonumber 
V= 1+ {\bf a}\cdot {\bf x} \;\;\; , \;\;\; W=1\;\;\;\;\;\;\mbox{and}\,\,\,\,\,F=V ,
\end{eqnarray}
one gets
\begin{equation}
\label{f13}
{\cal{H}}'' = \left[ {\atop} m + m \; {\bf a} \cdot
{\bf x} + \frac{\hat{{\bf{p}}}^2}{2m}+ \frac{1}{2m} \hat{{\bf{p}}}\cdot ({\bf a}\cdot {\bf x})\hat{{\bf{p}}} \right] \eta. \;\;\; 
\end{equation}
In (\ref{f12}) and (\ref{f13}) we have neglected the higher order
relativistic and gravitational/inertial terms.

For the particle $m$ far away from the body $M$ one can neglect the terms $\frac{3}{2m} \hat{{\bf{p}}} \cdot({\bf g} \cdot {\bf x}) \hat{{\bf{p}}}$\, and $\frac{1}{2m} \hat{{\bf{p}}}\cdot ({\bf a}\cdot {\bf x})\hat{{\bf{p}}}$ in (\ref{f12}) and (\ref{f13}), respectively, since they are less than the kinetic term by a factor of $GM/r \sim 10^{-6}$ (for the gravitational field of the Earth) and much weaker by several orders than the leading and next to leading order terms linear in $m$. In(\ref{f13}) we are assuming that ${\bf a}$ is such that $|{\bf a}\cdot {\bf x}| \sim GM/ r$. The Darwin term contributions in these expansions are zero in each case; in fact, in (\ref{f12}) we have $\nabla^2 F \,=\, 0$ (far away from the source and in the approximation considered) and in (\ref{f13}) for obvious reasons. Then, we come to the conclusion that the conformal coupling is in agreement with the
 equivalence principle.  

Last but no least, we call attention to the fact that we are not claiming that the conformal coupling is the correct coupling for the various scalar particles. The question of which value(s) of $\lambda$ should constitute the correct coupling to gravity depends on the particular field theory used for the scalar field (see, e.g. \cite{faraoni1} and references therein). Given the current theoretical situation it seems more of an experimental problem to identify which would be the correct $\lambda$ coupling(s) for the various scalar particles. 

\vspace{1cm}
 
\noindent {\bf Acknowledgements}

A.A. thanks CNPq-Brazil and H. B. to  FAPESP-Brazil for financial supports. The authors thank the referee for relevant comments and criticisms.


\begin{thebibliography}{**}
\bibitem{nota1} The operator $\theta$ enters the second order field equation (\ref{secondorder}), $\frac{\partial^2}{\partial t^2} \phi - \theta \phi = 0 $. Therefore the theory of elliptic diferential operators applies \cite{fulling}.
\bibitem{nota2} 
For charged spin-0 particles the first relativistic corrections to the Coulomb potential are of order $(p/m)^4$, whereas, in the spin $1/2$ case they appear in $(p/m)^2$ (spin-orbit and Darwin terms)\cite{feshbach}.
\end{thebibliography}
\end{document}